\documentclass[aps,prl,twocolumn,showpacs,showkeys,nofootinbib,superscriptaddress]{revtex4-1}

\usepackage[utf8]{inputenc}
\usepackage{graphicx}
\usepackage{amsfonts}
\usepackage{amssymb}
\usepackage{amsmath}
\usepackage[mathscr]{eucal}
\usepackage{setspace}
\usepackage{booktabs}
\usepackage[shellescape,latex]{gmp}
\usempxpackage{amssymb}

\def\sqr#1#2{{\vcenter{\hrule height.#2pt
   \hbox{\vrule width.#2pt height#1pt \kern#1pt
      \vrule width.#2pt}
   \hrule height.#2pt}}}

\def\bsqr#1#2{{\vrule width #1pt height#2pt}}
\def\bsquare{{\mathchoice\bsqr66\bsqr66\bsqr33\bsqr33}}
\def\badbreak{\penalty1000}

\def\fir{{\scriptscriptstyle{\text{\rm IR}}}}                 
\def\fuv{{\scriptscriptstyle{\text{\rm UV}}}}              
\def\lm0{{\lambda_0}}                                             
\def\nrN{N}                                                       
\def\v{b}                                                               

\def\leff{\ell}                                              
\def\Lmax{L_{\text{max}}}                                    
\def\lamA{\lambda_{\scriptscriptstyle{\text{\rm A}}}}        

\newcommand*{\GtrSim}{\smallrel\gtrsim}
\newcommand*{\LessSim}{\smallrel\lesssim}

\newcommand*{\LessApprox}{\smallrel\lessapprox}

\makeatletter
\newcommand*{\smallrel}[2][.8]{%
  \mathrel{\mathpalette{\smallrel@{#1}}{#2}}%
}
\newcommand*{\smallrel@}[3]{%
  \sbox0{$#2\vcenter{}$}%
  \dimen@=\ht0 %
  \raise\dimen@\hbox{%
    \scalebox{#1}{%
      \raise-\dimen@\hbox{$#2#3\m@th$}%
    }%
  }%
}
\makeatother

\usepackage{amsmath} 
\usepackage{dsfont}  
\usepackage{bm}      

\def\beq{\begin{equation}}
\def\eeq{\end{equation}}
\def\beqs#1\eeqs{\beq\begin{split} #1 \end{split}\eeq}

\long\def\comment#1{}

\usepackage{microtype}
\usepackage[dvipsnames]{xcolor}
\usepackage[colorlinks=true,backref=false, linktocpage=true,
citecolor=PineGreen,urlcolor=PineGreen,linkcolor=PineGreen,pdfpagemode=UseOutlines]{hyperref}

\hypersetup{%
  bookmarksnumbered=true,
  pdftitle = {},
  pdfsubject = {},
  pdfauthor = {},
  pdfkeywords = {}
}

\begin{document}

\title{Anderson Metal-to-Critical Transition in QCD}

\author{Andrei\ Alexandru}
\email{aalexan@gwu.edu}
\affiliation{The George Washington University, Washington, DC 20052, USA}

\author{Ivan Horv\'ath}
\email{ihorv2@g.uky.edu}
\affiliation{University of Kentucky, Lexington, KY 40506, USA}
\affiliation{Nuclear Physics Institute CAS, 25068 \v{R}e\v{z} (Prague), Czech Republic}

\date{Jun 12, 2022}

\begin{abstract}

A picture of thermal QCD phase change based on the analogy with metal-to-insulator 
transition of Anderson type was proposed in the past. In this picture, 
a low-$T$ thermal state is akin to a metal with deeply infrared (IR) Dirac 
modes abundant and extended, while a high-$T$ state is akin to an insulator 
with IR modes depleted and localized below a mobility edge $\lamA \!>\!0$. 
Here we argue that, while $\lamA$ exists in QCD, a high-$T$ state is not 
an insulator in such an analogy. Rather, it is a critical state arising due 
to a new singular mobility edge at $\lambda_\fir \!=\!0$. This new mobility 
edge appears upon the transition into the recently proposed IR phase.
As a key part of such a {\em metal-to-critical} scenario, we present evidence 
using pure-glue QCD that deeply infrared Dirac modes in the IR phase extend 
to arbitrarily long distances. This is consistent with our previous suggestion 
that the IR phase supports scale invariance in the infrared. We discuss 
the role of Anderson-like aspects in this thermal regime and emphasize that 
the combination of gauge field topology and disorder plays a key role in 
shaping its IR physics. Our conclusions are conveyed by the structure of Dirac 
spectral non-analyticities.

\medskip

\keywords{QCD phase transition, quark-gluon plasma, IR phase, Anderson localization, 
          scale invariance}
\end{abstract}

\maketitle


\noindent
{\bf 1.~Introduction. $\,$}
A fruitful aspect of progress in physics is the interplay of ideas between 
particle/nuclear and condensed matter areas. The development of 
the renormalization group~\cite{Fisher:1998kv,Kadanoff2013-KADRTV-2} or the complicated 
history of Anderson-Higgs mechanism~\cite{Quigg:2015cfa,Ellis:2012sy} 
illustrate this point. Among the ideas that originated in the realm 
of condensed matter, the notion of disorder-induced (Anderson) 
localization~\cite{Anderson:1958a} had a strong impact on other parts 
of physics. Yet, its influence on the basic framework of elementary particle 
physics (Standard Model) has not been substantial.

One context where Anderson localization may prove relevant is in thermal phase 
changes of strongly interacting matter. Here the electron eigenstates in random 
potentials of Anderson type could be analogous to Dirac quark modes in backgrounds 
of non-Abelian gauge fields at finite temperature. If the influence of pure disorder 
on electron dynamics turns out to be similar to that of thermal agitation on 
quarks, this could be important for understanding the mechanism of QCD transition.

A decade after it was first mentioned that an Anderson-like feature could appear 
in finite temperature Dirac spectra~\cite{Halasz:1995vd}, it was proposed in 
Refs.~\cite{GarciaGarcia:2005vj, GarciaGarcia:2006gr} that the QCD (chiral) 
transition may be a disorder-driven phenomenon analogous to Anderson 
metal-to-insulator transition. Here the ``insulating" nature of a hot thermal state 
expresses the absence of long-distance ($>\,$1/T) physics. 
It is specific to this scenario that, in addition to a depletion of the Dirac 
spectrum in this range, such physics is inhibited by any remaining infrared (IR) 
modes being exponentially localized at shorter scales.

The existence of Anderson-like features in hot QCD has been investigated
and the critical spectral point $\lamA \!>\! 0$ has been 
found~\cite{Kovacs:2010wx, Giordano:2013taa,Ujfalusi:2015nha}.
However, describing a hot thermal state as an Anderson insulator is at odds 
with its chiral polarization properties~\cite{Alexandru:2014paa} which
suggest the presence of long-distance physics. This inconsistency became 
sharp in light of our recent suggestion~\cite{Alexandru:2019gdm} that, above 
the crossover temperatures ($\approx\! 150\!-\!180\,$MeV), there is a thermal 
phase transition at temperature $200 < T_\fir < 250\,$MeV, involving 
a proliferation of deep IR modes ($0 \!\LessSim\! \lambda \!\ll\! T$), and 
the appearance of long-range scale-invariant IR physics. This new dynamical 
regime is referred to as the IR phase. 

Although metal-to-insulator scenario (no IR physics) is incompatible with 
IR phase (long-range physics), a fruitful Anderson-like interpretation can 
be given to Dirac modes of IR phase, while keeping the long-range feature 
intact. Clarifying this is the main focus of the present paper.

Few contextual details are important before we start. 
{\em(1)} A fixed 3d Anderson model can describe varied electronic behaviors 
(conductor, insulator, criticality) simply by suitably choosing the Fermi 
level. This freedom does not exist in QCD, whose symmetries require such
reference to be at zero. Hence, the Anderson analog of the QCD thermal state, 
if any, is determined by properties of modes near $\lambda_\fir \!\equiv\!0$.
{\em (2)} Despite its IR mode proliferation, a thermal state in the IR phase 
could be an insulator if these modes were exponentially localized 
in IR spectral regime $\lambda \!<\! T$ with bounded range. A key ingredient 
in reconciling IR phase with Anderson-like aspects will be the evidence that 
this does not occur. Rather, we conclude that the length scale associated 
with modes at eigenvalue $\lambda$ diverges for $\lambda \!\to\! 0$.
{\em (3)} The available evidence is consistent with the mobility edge $\lamA$ 
appearing upon crossing into the IR phase. This point is solid in pure-glue 
QCD~\cite{Alexandru:2021pap} and, although existing full QCD estimates are 
somewhat below the $T_\fir$ window~\cite{GarciaGarcia:2006gr,Kovacs:2012zq}, 
a reliable determination is lacking. [We thank T.~Kov\'acs for clarifying 
this.] In either case, thermal state at $T \!<\! T_\fir$ would not act as 
an insulator since deep IR modes here are at least as mobile as in IR phase 
(see above).

Item {\em (2)} above points to the presence of singular mobility edge
at $\lambda_\fir \!=\!0$. Indeed, the length scale $\leff$ we will use to 
characterize modes is a proxy to exponential localization range: if modes
exponentially decay in space, their $\leff$ is related to localization 
range by a finite transformation with their ratio approaching unity at 
large $\leff$. Our approach to show {\em (2)} will follow the one we used 
to study spatial dimensions of modes in the vicinity of 
$\lambda_\fir$~\cite{Alexandru:2021pap}. To recapitulate, exact zeromodes 
were found to be $d_\fir \!=\! 3$ whereas the deep IR modes have lower 
dimensions, with $d_\fir \!\to\! 2$ for $\lambda \!\to\! 0$.
This makes the isolated point $\lambda \!=\! \lambda_\fir$ similar to 
the band of extended modes $\lambda \!\ge\! \lamA$, and deep IR modes 
($\lambda \GtrSim \lambda_\fir$) akin to those near $\lamA$ 
($\lambda \LessSim \lamA$). Here we find similar structure in metric 
properties which, among other things, leads to {\em (2)}. 
Hence, the deep-IR Dirac dynamics of QCD in the IR phase is similar 
to Anderson-like critical dynamics: rather than metal-to-insulator, 
the relevant analogy for QCD phase transformation occurring at 
$T_\fir$ is that of a {\em metal-to-critical} transition.



\smallskip
\noindent
{\bf 2. QCD Perspective.} 
This work is part of our ongoing~program to infer the properties of 
QCD vacuum and thermal states from spectral features of the Dirac 
operator. At the lattice-regularized level, the probing Dirac operator 
can be chosen to best reflect the properties of interest, with 
universality expected in the continuum limit. Given the important
role of quark chirality and glue topology in low-energy QCD, we use 
the overlap operator~\cite{Neuberger:1997fp} for these purposes. 
Refs.~\cite{Alexandru:2012sd,Alexandru:2014paa,Alexandru:2015fxa,
Alexandru:2019gdm,Alexandru:2021pap} are most relevant to 
the topic studied here.

The overlap spectral feature from which the notion of IR phase 
eventually arose, the infrared mode density peak, was first 
observed in Ref.~\cite{Edwards:1999zm}. Renewed interest appeared 
when its peculiar chiral polarization properties were 
noted~\cite{Alexandru:2014paa}. Further studies showed that 
these IR modes persist both in the thermodynamic and 
the continuum limit~\cite{Alexandru:2015fxa}. This 
led to a new perspective in which the peak reflects a dynamical 
IR property of glue that drives the phase changes in SU(3) gauge theories 
with fundamental quarks~\cite{Alexandru:2015fxa}. 
We later identified this property with IR scale invariance 
of strongly coupled glue fields~\cite{Alexandru:2019gdm}.

The relevance of Anderson-Dirac feature at $\lamA$ in this 
context was pointed out in Ref.~\cite{Alexandru:2021pap}. Remarkably,
such spectral non-analyticity is exactly what is needed to realize 
the IR-UV decoupling important for the proposed mechanism of 
IR scale invariance~\cite{Alexandru:2019gdm}. This non-analyticity
is manifest in dimension $d_\fir(\lambda)$ of the modes, whose 
structure near $\lambda_\fir \!=\!0$ was found to be similar to
that near $\lamA$. Here we study 
the spectral function $\gamma(\lambda)$ expressing the ``sizes" 
(distances) of the modes and find the same qualitative scenario. 
We thus propose that an Anderson-like aspect is present 
not only at $\lamA$ but also at $\lambda_\fir$. 
Hence, although the origin of IR phase is likely topological and 
complex~\cite{Alexandru:2021pap}, some of its IR properties, 
including certain features of scale invariance, may be understood 
in terms of Anderson-like criticality.

\smallskip
\noindent
{\bf 3. The $\gamma$-index.} 
We study thermal QCD represented by a field system in 4-volume $L^3/T$ 
and regularized via hypercubic lattice with shortest distance $a$. Scales 
$1/L$ and $1/a$ represent the IR and UV cutoffs. Our analysis will 
focus on eigenmodes $D \psi_\lambda(x) \!=\! i \lambda \psi_\lambda(x)$ 
of Dirac operator $D$, represented on the lattice by the overlap 
operator.

Unlike the effective volume whose definition requires careful 
consideration~\cite{Horvath:2018aap,Horvath:2018xgx}, an acceptable effective 
distance spanned by a mode can be defined as a simple probability mean. 
Indeed, let $\psi \!=\! \psi(x_i)$ be a generic eigenmode with 
$x_i$, $i=1,2,\ldots,\nrN = (L/a)^3/(Ta)$, denoting the positions of 
lattice sites. The probability vector $P$ associated with $\psi$ is specified 
by $P \!=\! (p_1,p_2,\ldots,p_\nrN)$, $p_i = \psi^+ \psi(x_i)$. The effective 
distance can be quantified by the average separation from the location
of the probability maximum, namely
\begin{equation}
   \leff[\psi] = \leff[P] = \sum_{i =1}^\nrN \, p_i \, | x_m - x_i |
   \quad , \quad  
   p_m \ge p_i \;,\; \forall \,i
   \label{eq:030}                    
\end{equation}
where $| \ldots |$ is the Euclidean norm on a periodic lattice.
Distance $\leff[\psi]$ can be viewed as a version of 
``mode radius". Its response to the removal of IR cutoff
\begin{equation}
   \bigl\langle \; \leff[\psi] \; \bigr\rangle_{a,L,\lambda} 
   \,\propto\,\, L^{\gamma(a,\lambda)} 
   \quad \text{for} \quad L \to \infty  \quad 
   \label{eq:060}	
\end{equation} 
defines the index $0 \le \gamma \le 1$. Note that $\gamma \!=\! 0$ for 
exponentially localized states and $\gamma \!=\! 1$ for plane 
waves.

\begin{figure}[t]
\begin{center}
    \vskip -0.10in
    \centerline{
    \hskip 0.0in
    \includegraphics[width=8.0truecm,angle=0]{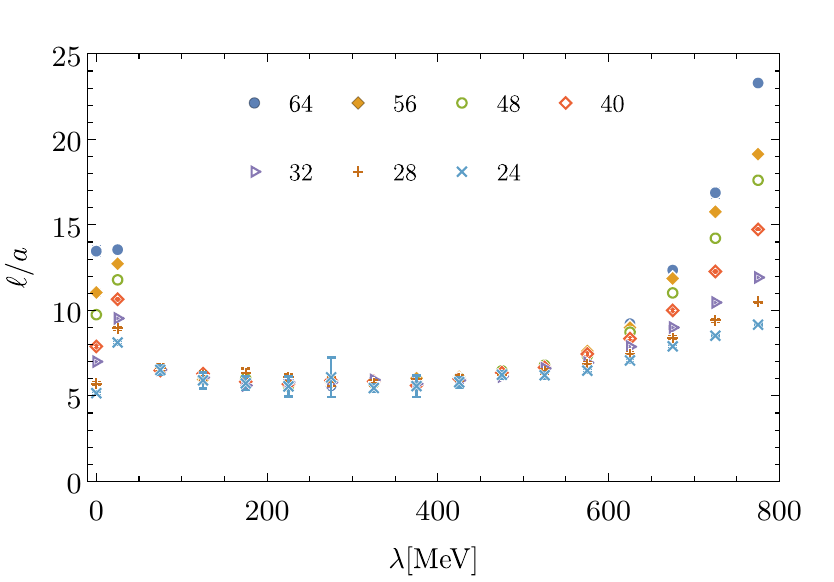}
    }
    \vskip -0.08in
    \caption{The spectral dependence of effective distance $\leff$ 
    for Dirac modes at various IR cutoffs. The legend specifies $L/a$.}
    \label{fig:ell_vs_lam}
    \vskip -0.45in
\end{center}
\end{figure}

\smallskip
\noindent
{\bf 4. The Results.} 
We studied the above characteristics in pure-glue lattice QCD at 
$T\!=\!1.12\, T_\fir$. Wilson action with 
$\beta \!=\! 6.054$ ($a\!=\!0.085\,$fm, $r_0\!=\!0.5\,$fm) was 
used, and the eigenmodes of the overlap operator ($\rho \!=\! 26/19$)
were analyzed on systems with
$L/a \!=\!24,28,32,40,48,56,64$ and $1/(Ta)\!=\!7$. 
The implementation of overlap is described 
in Refs.~\cite{Alexandru:2011sc, Alexandru:2014mqy, Alexandru:2011ee}.
Additional technical details can be found 
in Refs.~\cite{Alexandru:2019gdm, Alexandru:2021pap}.
Overlap discretization is crucial to distinguish zero modes from 
the nearby modes. 
On our largest volumes the low-lying modes have unusually small 
eigenvalues and we numerically diagonalize the overlap operator $D$
rather than more conventional $D^\dagger D$. Using this approach,
we find that low-lying modes are separated from zero modes by 
several orders of magnitude. 


Our basic finding is already conveyed by coarse spectral data 
for $\leff$ shown in Fig.~\ref{fig:ell_vs_lam}. 
We observe a central plateau, insensitive to the IR cutoff, 
and the ``rises" near $\lambda_\fir \!=\!0$ and 
$\lamA \!\approx\! 750\,$MeV, where $\ell$ increases with $L$. 
Point $\lamA$, marking the Anderson-like critical point, was 
determined for this setup in Ref.~\cite{Alexandru:2021pap}. 
The two IR-responsive regimes act as
``finite-volume critical regions" in Anderson-like language.
The novelty is the existence of one near $\lambda_\fir$.


\begin{figure}[t]
\begin{center}
    \vskip -0.10in
    \centerline{
    \hskip 0.0in
    \includegraphics[width=8.4truecm,angle=0]{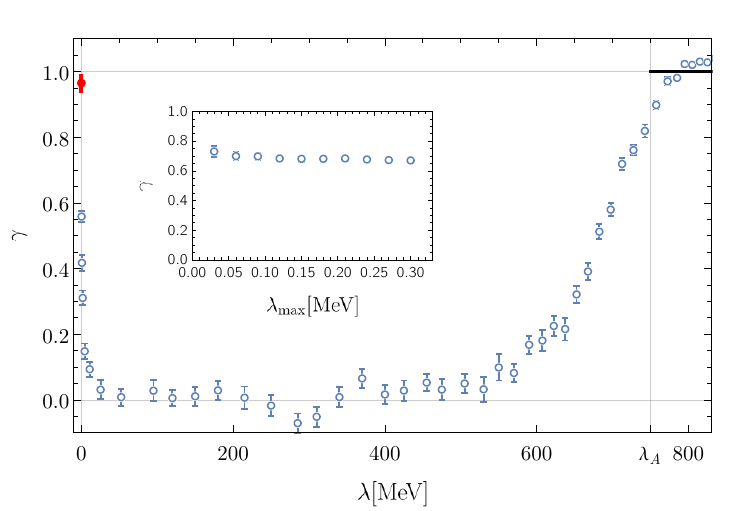}
    }
    \vskip -0.08in
    \caption{The spectral dependence of index $\gamma$ in pure-glue QCD.}
    \label{fig:gamma_vs_lam}
    \vskip -0.40in
\end{center}
\end{figure}

\begin{figure}[t]
\begin{center}
    \vskip -0.1in
    \centerline{
    \hskip 0.00in
    \includegraphics[width=8.4truecm,angle=0]{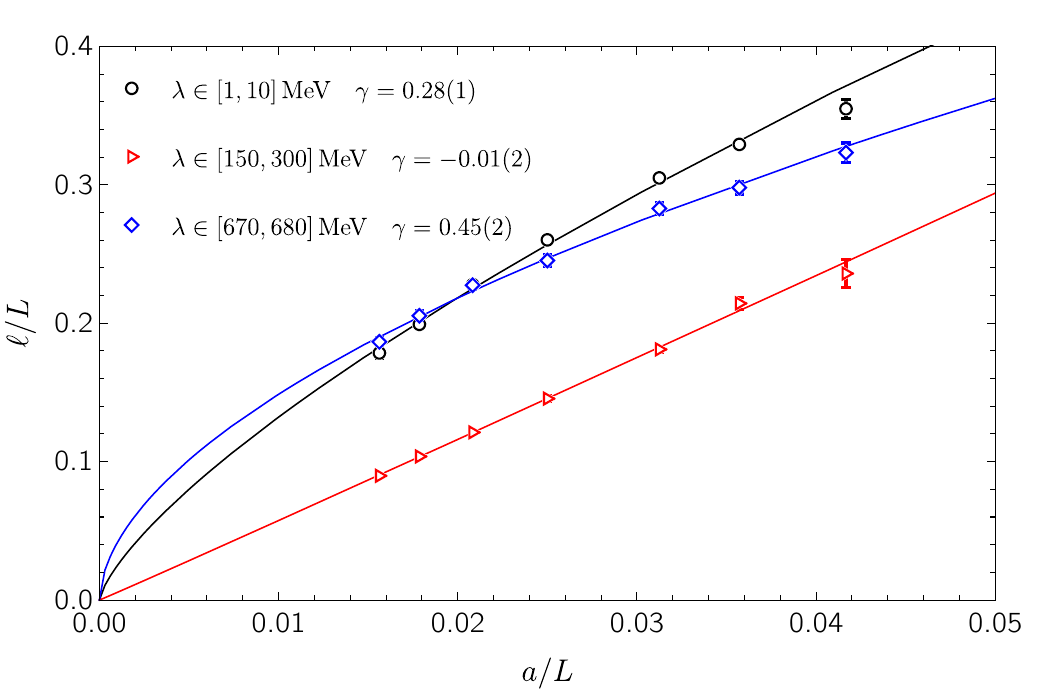}
    }
    \vskip -0.05in
    \caption{Examples of scaling in $\ell/L$ within the plateau (triangles), 
    left rise (circles) and right rise (diamonds).}
    \label{fig:examp_scaling}
    \vskip -0.30in
\end{center}
\end{figure}

Asymptotic tendency to extend into distance is expressed by 
index $\gamma$, Eq.~\eqref{eq:060}. Numerically, we compute this 
index by fitting $\leff$, obtained on different ensembles, 
to a power-law in $L$. The errors on $\gamma$ are calculated using 
standard $\chi^2$-fitting methods.
Its spectral behavior, shown in Fig.~\ref{fig:gamma_vs_lam}, 
characterizes the above features of raw data in a robust manner.
Thus, we again observe a plateau, approximately centered around 
the scale of temperature, with $\gamma \approx 0$. The latter 
is consistent with the presumed exponentially localized nature 
of these modes. At the same time, the left and right rises appear 
below $\lambda \approx 50\,$MeV and above $\lambda \approx 550\,$MeV
respectively. 
The examples illustrating the scaling behavior of $\ell/L$ in these
three spectral regimes are shown in Fig.~\ref{fig:examp_scaling}.
The red point in Fig.~\ref{fig:gamma_vs_lam} is the finite volume
result for zero modes. Extrapolated~value $\gamma \!=\! 1$ is 
predicted by their $d_\fir \!=\! 3$ 
\cite{Alexandru:2021pap}, and has indeed materialized in the present 
$\ell[\psi]$ data. The black horizontal line above $\lamA$ marks 
the extended regime of the bulk. 

A notable feature of our results is conveyed by the inset of
Fig.~\ref{fig:gamma_vs_lam}, showing $\gamma$ for near-zero 
modes from variable interval $(0,\lambda_{\text{max}})$. Data 
from available volumes suggest that 
$\lim_{\lambda \to \lambda_\fir^+} \gamma(\lambda) \!\LessApprox\! 0.8$.
The left vicinity of $\lamA$ (vertical line in 
Fig.~\ref{fig:gamma_vs_lam}) also suggests $\gamma \!<\!1$. 
We thus observe a discrepancy of these limiting values relative 
to both fully extended ($\gamma \!=\! 1$) and exponentially 
localized ($\gamma \!=\! 0$) regimes. This suggests that 
$\gamma(\lambda)$ has the same kind of critical non-analyticity 
structure as mode dimension $d_\fir(\lambda)$. 
(See Fig.~5 in Ref.~\cite{Alexandru:2021pap}.)

\begin{figure}[b]
\begin{center}
    \vskip -0.05in
    \centerline{
    \hskip -0.02in
    \includegraphics[width=4.5truecm,angle=0]{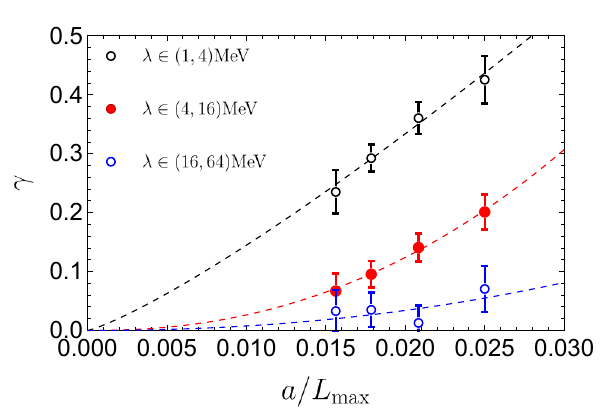}
    \hskip -0.12in
    \includegraphics[width=4.5truecm,angle=0]{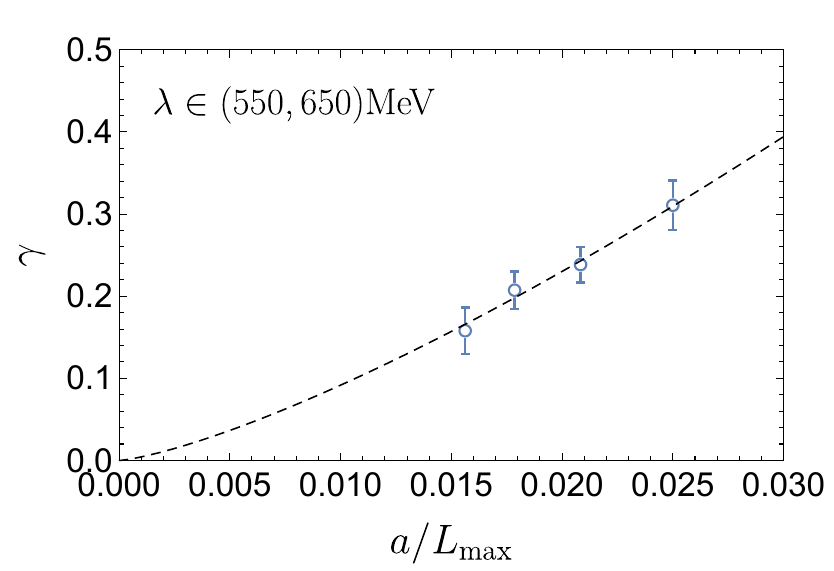}    
    }
    \vskip -0.05in
    \centerline{
    \hskip -0.02in
    \includegraphics[width=4.5truecm,angle=0]{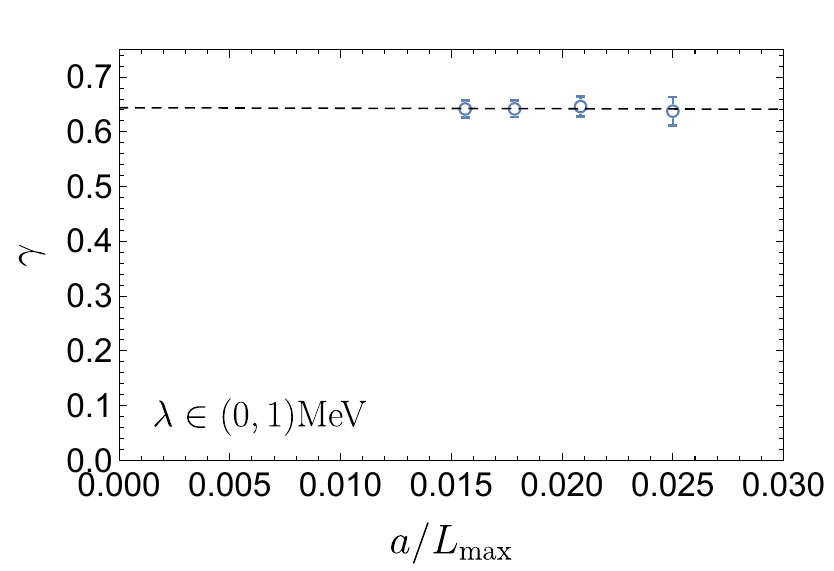}
    \hskip -0.12in
    \includegraphics[width=4.5truecm,angle=0]{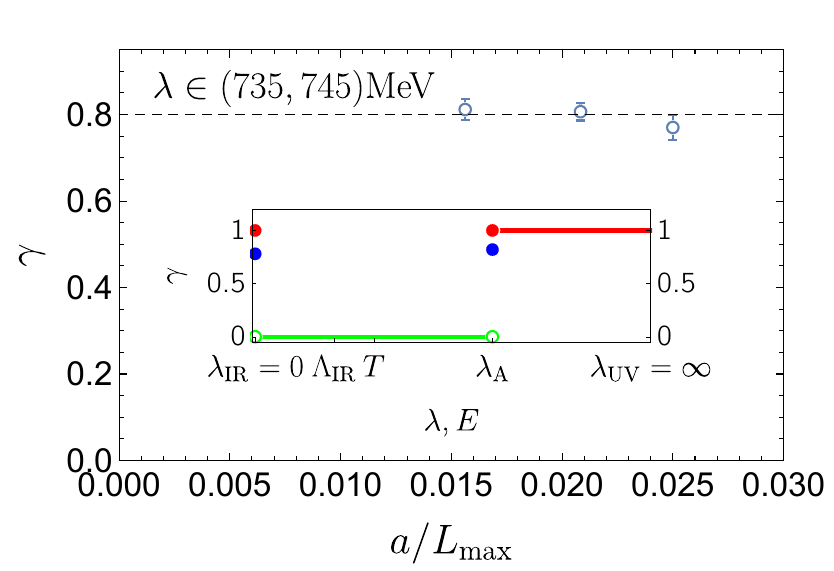}    
    }
    \vskip -0.07in
    \caption{Index $\gamma$ as a function of $1/\Lmax$ within left/right rise 
    (upper left/right) and very near $\lambda_\fir$/$\lamA$ (lower left/right). 
    The extrapolations use the power (top) and constant~(bottom) fits.}
    \label{fig:slide_rises}
    \vskip -0.40in
\end{center}
\end{figure}

Such scenario is supported by volume tendencies in $\gamma$. 
Fig.~\ref{fig:slide_rises} shows the dependence of $\gamma$ 
on $1/\Lmax$ within the two rises, with $\Lmax$ being the maximal 
lattice size used in the analysis. Four systems of consecutive 
available sizes were used to extract each value shown from 
a power fit. The top plots suggest the extrapolation 
toward $\gamma \!=\!0$ for interiors of the rises, 
while in the immediate vicinity of $\lambda_\fir$ and $\lamA$ 
(bottom plots) the values remain stable.
In case of $\lambda_\fir$ we split the outer critical region into 
three parts with approximately equal statistics of modes to show
that $\gamma \!\to\! 0$ in each.
This indeed supports the singularity structure shown in the inset 
of the bottom right plot.

\smallskip

\noindent
{\bf 5. Discussion.} 
Important consequence of our findings is that the length scale $\ell$ 
of non-zero modes in IR regime $0 \!<\! \lambda \!<\! T$ is unbounded 
in infinite volume. Indeed, $\ell(\lambda)$ increases toward infrared 
($\lambda \to 0^+$, left rise). Its rate of growth with $L$, namely
$\gamma(\lambda)$ increases toward infrared as well 
(Fig.~\ref{fig:gamma_vs_lam}), with deepest near-zero modes governed 
by $\gamma(0^+) \!\LessApprox\! 0.8$
(inset of Fig.~\ref{fig:gamma_vs_lam}). Hence, $\ell(\lambda)$ in 
infinite volume diverges, at least for $\lambda \!\to\! 0^+$. Since 
$\gamma(\lambda,L_{\text{max}})$ tends to zero with $L_{\text{max}}$, 
at least for $\lambda \!>\! 1\,$MeV (Fig.~\ref{fig:slide_rises}, top
left), we conclude that the most likely scenario is that 
$\ell(\lambda)$ 
remains finite at any fixed $0 \!<\! \lambda \!<\! T$, and 
$\lambda_\fir \!=\! 0$ is a new mobility edge. Given the point {\em (1)} 
of Sec.~1, this in turn means that crossing into IR phase can be 
described as {\em metal-to-critical} transition.

In Ref.~\cite{Alexandru:2021pap} we proposed a new feature of QCD 
Dirac modes in IR phase: their tendency to spread into available volume, 
expressed by IR dimension function $d_\fir(\lambda)$, is
non-analytic at $\lambda_\fir \!=\! 0$ and $\lamA \!>\! 0$. 
Both non-analyticities feature a characteristic double discontinuity. 
Results of the present work suggest a similar structure in mode tendency 
to extend into available distances $\gamma(\lambda)$ 
(inset of Fig.~\ref{fig:slide_rises}). Since $d_\fir$ and $\gamma$ 
reflect distinct geometric features of Dirac modes (volume and distance), 
the consistency of these findings reinforce the picture of $\lambda_\fir$ 
and $\lambda_A$ as two points of Dirac non-analyticity present in IR phase.

The chief ramification of the above is that the vicinity of $\lambda_\fir$ 
can support long-range physics. Indeed, $\ell(\lambda) \!\ge\! 1/T$ for
all $\lambda < T$, and diverges for $\lambda \!\to\! 0^+$. This is 
a prerequisite for IR phase to support IR scale-invariant 
glue~\cite{Alexandru:2019gdm}. Moreover, the added aspect of Anderson-like 
criticality will likely play a relevant role in the eventual full 
understanding of the scale invariant dynamics involved. More concretely, 
we expect the modes in IR ``critical region" of size $1/L$ to encode 
the zero glueball screening masses conjectured to characterize 
the IR phase~\cite{Alexandru:2019gdm}.  


The presented evidence is in the context of pure-glue QCD and the analysis 
should be extended to QCD with physical quarks. We point out, however, 
that the overlap-inferred glue response to IR transition has been 
previously found to be very similar in these 
theories~\cite{Alexandru:2019gdm}.

\begin{figure}[t]
\begin{center}
    \vskip -0.10in
    \centerline{
    \hskip 0.00in
    \includegraphics[width=8.2truecm,angle=0]{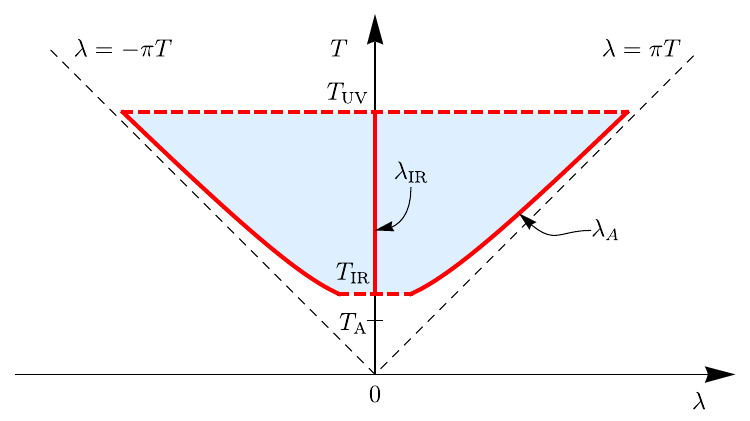}
    }
    \vskip -0.05in
    \caption{Conjectured Dirac spectral phase diagram of thermal pure glue QCD.}
    \label{fig:Dirac_phases}
    \vskip -0.40in
\end{center}
\end{figure}

The present results combined with those of 
Refs.~\cite{Alexandru:2019gdm, Alexandru:2021pap} suggest a Dirac 
spectral ``phase diagram" of thermal QCD similar to one shown in 
Fig.~\ref{fig:Dirac_phases}. The thick red lines are phase boundaries 
separating the regular $d_\fir \!=\! 3$,  $\gamma \!=\! 1$ modes 
outside of the closed region, from the rest. Solid parts of 
the boundary feature the double discontinuity structure. 
In the absence of evidence to the contrary, the dashed parts 
are presumed to be simple discontinuities.
Critical lines $\pm \lamA(T)$ coincide~\cite{Alexandru:2021pap} 
with mobility edges studied previously~\cite{Kovacs:2010wx, 
Giordano:2013taa, Ujfalusi:2015nha}.
The critical line $\lambda_\fir(T) \!=\! 0$ is also the line of 
singularities in mode density $\rho(\lambda,T) \!\propto\! 1/\lambda$ 
for $\lambda\to 0$, defining the IR phase~\cite{Alexandru:2019gdm}. 
Note that the non-zero value of $\lamA(T_\fir)$ 
(see also~\cite{Kov2021_A}) allows for the possibility that the two 
types of critical lines have distinct critical properties.

Finally, we wish to make several comments. 
{\bf (i)} 
The range $(-\lamA,\lamA)$ widens with increasing $T$ due to 
growing disorder.
However, increasing $T$ also weakens the interaction which biases 
modes toward free-like mobility and hence softens the effect of 
disorder alone. In fact, it is unknown how these competing tendencies 
resolve near the expected onset $T_\fuv$ of perturbative plasma, 
where critical lines end. Indeed, it would also be consistent with 
asymptotic freedom if lines $\pm\lamA(T)$ turned back toward 
the $T$-axis at sufficiently high temperatures, but there is currently 
no evidence of that. We also note that the possibility that
$T_\fuv \to \infty$ cannot be excluded at this point.
{\bf (ii)} 
It would be valuable to understand the full scope and limitations 
of similarities between 3d Anderson transitions and QCD points 
$\lamA$, $\lambda_\fir$. This is non-trivial since,
although the former are simple idealizations describing 
the effects of pure disorder on free particle propagation, 
the latter express the influence of thermal disorder on 
structured and complicated dynamics. Nevertheless, the existence 
of two distinct critical lines in the QCD spectral diagram offers 
a schematic understanding of this added complexity. Indeed, 
$\lamA(T)$ can be understood at high temperatures 
$T \LessApprox T_\fuv$ 
as resulting from the above competition of disorder and asymptotic 
freedom. The appearance of $\lambda_\fir$, on the other hand, 
is natural when $T$ approaches $T_\fir$ from below. 
Here the increased thermal fluctuations erode IR non-perturbative 
fields, until only those essential for supporting glue topology 
remain, being the most impervious to disorder. This is 
the nature of the transformation occurring at $T_\fir$.
The resulting prominent role of glue topology in the IR phase is 
supported by integer effective dimensions~\cite{Alexandru:2021pap} 
and other results~\cite{Vig:2021oyt,Cardinali:2021mfh}.
{\bf (iii)} 
Incorporating the above into the original IR phase 
proposal~\cite{Alexandru:2019gdm} suggests the following scenario. 
As the temperature of the system increases, 
the Dirac spectral crossover at $T_A \!\approx\! 150\,$MeV 
signals the onset of condensate melting. 
By the time the temperature $T_\fir$ ($T_\fir \in (200, 250)\,$MeV) 
is reached, thermal fluctuations have eroded away most non-perturbative 
IR fields, including those responsible for the scale anomaly. 
At this point a phase 
transformation occurs in which the system separates into the scale 
invariant IR component, characterized by the power-law mode 
density $\rho(\lambda,T) \!\propto\! 1/\lambda$, and the bulk. 
This is accompanied by the appearance of Anderson-like critical 
lines $\lambda_\fir(T) \!=\! 0$ 
and $\pm\lamA(T)$ (Fig.~\ref{fig:Dirac_phases}). 
The critical region associated with $\lambda_\fir$ gives 
the physics of the IR phase its long-range character and the
attribute of competition between glue topology and disorder. 
The criticality of $\lamA$, on the other hand, provides a 
localization barrier that makes the IR component independent 
of the bulk, effectively allowing for non-analytic running and 
scale invariance below some energy scale $\Lambda_\fir \!<\! T$. 
The two-component nature of IR phase and its IR scale invariance 
are expected to be important in generating the near-perfect 
fluidity of hot strongly interacting matter.
{\bf (iv)} 
The features described in this work are also relevant for chiral
limit considerations. Here the problem of $U_A(1)$ anomaly survival 
at high temperatures~\cite{Pisarski_1984} is an open question. 
For recent complementary views on this see, e.g., 
Refs.~\cite{Ding:2020xlj,Aoki:2020noz,Kaczmarek:2021ser}. 
One should also mention here that the validity of the original 
metal-to-insulator scenario in the chiral limit remains an open 
question in a similar sense.


\begin{acknowledgments}
A.A. is supported in part by the U.S. DOE Grant No. DE-FG02-95ER40907.
I.H. acknowledges the discussions with Peter Marko\v{s} and Robert Mendris. 
\end{acknowledgments}


\bibliography{my-references}

\end{document}